\documentclass{article}

\begin{document}

{\bf A POINTLESS MODEL FOR THE CONTINUUM AS THE FOUNDATION FOR QUANTUM
  GRAVITY}

\bigskip

by: 

{\bf Louis Crane, 

Mathematics Department 

Kansas State University,

Manhattan KS 

66502}

\bigskip

crane@math.ksu.edu

\bigskip

{\bf DATE:} March 28, 2008

\bigskip

{\bf SUMMARY} In this paper, we outline a new approach to quantum
gravity; describing states for a bounded region of spacetime as
eigenstates for two classes of physically plausible gedanken
experiments. We end up with two complementary descriptions in which
the point set continuum disappears.

The first replaces the continuum of events with a handlebody
decomposition of loop space. We conjecture that techniques fron
algebraic topology will allow us to extend state sum models on
spacetime to loop space.

The second picture replaces the continuum with a nondistributive
lattice; the classical limit seems more tractible in this picture.

\newpage

{\bf A POINTLESS MODEL FOR THE CONTINUUM AS THE FOUNDATION FOR QUANTUM
  GRAVITY}

\bigskip

by: Louis Crane, Mathematics Department KSU

\bigskip

We want to implement the recent suggestion [1] that the continuum
picture disappears in highly curved regions in quantum general relativity.

A body of recent work in relativity has shown
us that only a finite dimensional Hilbert space of information can
pass from any bounded region in spacetime to its exterior [2]. The
description of the geometry of a region as a
metric on an open subset of a manifold contains more information than
external observers can  ever detect. Such information can play no role in a
quantum theory of the observable geometry of the region.

Here we will outline an approach to the
construction of a quantum theory of the
geometry of a bounded region R in spacetime, using only
the results of external experiments probing the geometry of the
region. We assume the area of the spatial boundary of the R
is known, and that the exterior of R is simple, causal, flat,
topologically trivial, and can
be treated classically. This will produce
a relational geometry, since the probes would communicate 
different information to different external observers, and a pointless
geometry in that the finite information observable would not
distinguish an infinite point set.

We need to include the back reaction of the
probes on the spacetime geometry of R, which would focus 
the trajectories of
the probes and redshift the information, and cause the information
limitation mentioned above. Too large of a probe or too many probes
would cause an event horizon to form and therefore not transmit
information to the outside. Including the back reaction is in the
spirit of quantum mechanics, since we cannot ignore the effect of the
probe on the system, i.e. on the spacetime geometry of R.

We shall discuss two different kinds of probes, external, and
internal, which will give us two
complementary descriptions of R. Measuring one
type would disturb the result of the other, much as the
position and momentum
descriptions of a particle are complementary.

\bigskip

{\bf EXTERNAL PROBES. THE TELESCOPE-MICROSCOPE HYPOTHESIS }

\bigskip

In the first type of probing experiment, we imagine sending a probe
along some path through the region of spacetime. The probe is thought
of as including gyroscopes and a clock. This probe would measure the
holonomy of the connection on the spacetime along the path. In
a classical spacetime, assuming that measurements did
not disturb the spacetime geometry, we could imagine measuring holonomy
along each causal path, from which the metric could be recovered.

Quantum mechanically, however, we cannot think of a probe as following
a definite path, and the back reaction compresses the infinite number
of holonomies of the causal paths to a finite dimensional hilbert
space of information. 

\bigskip

At this point we propose:

\bigskip

{\bf TELESCOPE-MICROSCOPE HYPOTHESIS} {\it all the observable
information from causal path holonomies is contained in the families of
null geodesics from events in the causal past of R  to events in its
causal future.}

\bigskip 

The motivation for this suggestion is the thought that an
electromagnetic wave through R samples all
causal paths at once as in a path integral. Classically, a light pulse
would travel along null geodesics only because of Fermat's
principle. So we are proposing that the information which survives is
exactly the apparent position of all the classical images of past
events seen through R.

In making this hypothesis, we reason by analogy with the results of
astronomical observation of distant objects through highly curved
regions such as galaxies We are motivated in what follows by the
mathematical theory of gravitational lensing [3]. 

To restate, we replace the probes through all possible
causal paths with an optical gravitational lensing experiment, and conjecture
that the probabilities of the different multiple images exhaust the
transmissible information.

This hypothesis seems unnatural
for a large flat region of spacetime, but if we think of a region small
enough that quantum gravity is relevant, it would be full of
geometrical fluctuations of near black hole strength at near Planck
scale dimensions, so a pulse of light passing through the region would
have infinitely many multiple images as viewed by any observer in the future of R. 

The case of a large flat region could be included in this picture by
thinking of the wave optical corrections to a lensing experiment as
given by the average of all the multiple images in the quantum
fluctuations of the
geometry on the region, reproducing a path integral version of
quantum mechanics. This implements the suggestion that quantum general
relativity automatically imposes quantum mechanics on a particle [4].
Recovering standard quantum mechanical propagators from this picture will be a
stern test for this proposal.

\bigskip

{\bf MORSE THEORY ON LOOP SPACE}

\bigskip
Now let us consider the images of a wavefront generated at a single
event in the causal past of R, as seen by an observer in the future of R.

Since the elapsed time is a morse function on the space of paths
between two events, and since that space is homotopy equivalent to the
based loop space on R, $ \Omega (R)$,
the multiple images seen by each observer 
can be understood as
critical points of a morse function on $ \Omega (R)$. This is
equivalent to a handlebody decomposition of the loop space [5].

(If we combine the images all possible future observers could see, we
obtain a decomposition of the free path space of R. This is a
(contractible) complex of a slightly more
general type, since singularities of a higher order would generically
occur. Nevertheless, a type of cellular complex would result.)

In this description, the events or
localized subregions of R have disappeared, to be replaced by a geometry on
a cellular complex..
  
Next we need a theory that attaches probability
amplitudes to finite handlebody approximations to   $ \Omega (R)$,
and to the apparent sizes of the handles
to external observers (evidenced by the apparent distances between
images and the angular region surrounding R in which they
are observable.)

This problem is similar to the constructions of TQFTs  and
combinatorial state sums for quantum gravity in 3 and 4
dimensions [6,7]; because in homotopy theory  $ \Omega (R)$ looks
like a 3-manifold, while the differential graded algebra of complexes
on it can be obtained from that of R by the cobar construction, and
takes the form of a limit of multiple copies of the DG algebra on R [8].

If we replace R by a superposition of quantum amplitudes on
handlebodies, how could the classical picture of a spacetime region be
recovered?

\bigskip

{\bf INTERNAL PROBES}

\bigskip

Now let us imagine probing R by sending probes inside it which set off
pulses of light at some time. If R were a classical spacetime region,
we could reconstruct its spacetime geometry by noticing when and from
what angle different
external observers saw the same pulse, 
combining the past light cones of each
observer along its worldline into regions of Minkowski space, and
using the correlations for different observers to glue together the
patches of Minkowski space into a manifold. (In order to directly
observe the distance of an event, we should think of our observer as
binocular, and use parallax).

Different metrics on R would cause different identifications of
the apparent past patches, so the set of apparent positions for one
metric would not coincide with those for another.

If we take it that ``position'' in R means apparent position as
seen by all observers, then the set of apparent positions in R for a
{\it quantum} state of its geometry would not be described as subsets of a
point set. As we have argued in another paper [9], it would be a
non-distributive lattice, where the number of regions we could see
would be limited by the dimension of the hilbert space of
transmissible information, or the boundary area of R in Planck units.

In situations that were essentially classical, the nondistributivity
of the lattice would only appear at negligible scales, and the lattice
would be fine enough to approximate a continuous point set.

\bigskip

{\bf SUMMARY}

\bigskip

Our proposal reduces to the following steps:

\bigskip

1. Lift the BC model [7] for quantum gravity to  $ \Omega (R)$ using the
cobar construction,

\bigskip

2. Find the form of the transformation from the external probe basis,
as described by handlebody decompositions of  $ \Omega (R)$ to the
internal probe basis, described by non-distributive lattices of
apparent regions.

\bigskip

3. Investigate the form of the quantum theory in the internal probe
basis, and verify the classical limit.

\bigskip

This is conceptually similar to ordinary quantum mechanics, in which
dynamics are worked out in the momentum basis, then transformed to the
position basis to obtain the usual propagators, differential equations etc.

\bigskip

{\bf BIBLIOGRAPHY}

\bigskip

1. A. Ashtekar, M. Bojowald
Black hole evaporation: A paradigm,
Class. Quant. Grav. 22 (2005) 3349-3362
\bigskip

2. L. Susskind, An introduction to black holes, information and the string theory
revolution : 
the holographic universe, Singapore ; Hackensack, NJ : World Scientific, 2005.

\bigskip

3. V. Perlick, Gravitational Lensing from a Spacetime Perspective,
Living Reviews in Relativity, 2004 - emis.ams.org

\bigskip

4. S. Frittelli , C. N. Kozameh , E. T. Newman , C. Rovelli , R. S. Tate, 
Fuzzy spacetime from a null-surface version of GR,
Class.Quant.Grav. 14 (1997) 
A143-A154

\bigskip

5. J. Milnor, Morse Theory, Annals of Mathematics Studies, PUP 1969

\bigskip
6. L. Crane, L.H. Kauffman, D.N. Yetter, State Sum Invariants of 4-Manifolds
I, Arxiv preprint hep-th/9409167
\bigskip

7. J.W. Barrett, L. Crane,	A Lorentzian signature model for quantum
general relativity,
Class. Quantum Grav. 17 (2000) 3101-3118. 

\bigskip

8. J. McCleary, Users Guide to Spectral Sequences publish or Perish,
1985

\bigskip

9. L. Crane, What is the Mathematical Structure of Quantum Spacetime?, arXiv:0706.4452

\end{document}